\documentclass{article}

\usepackage{emulateapj}
\usepackage{graphics}

\def\linebreak{\hfil\break}

\def\
{\hfil\linebreak}
\def\singlespace{\baselineskip=14pt}

\def\singlespace{%
    \lineskip                .15ex
    \baselineskip            3.0ex
   \lineskiplimit              0ex
   \parskip                0.60ex plus .30ex minus .15ex
   }%

\def\etal{{\it et al}. }
\def\degree{\ifmmode {^\circ}\else {$^\circ$}\fi}
\def\mum{\ifmmode {\rm \mu {\rm m}}\else $\rm \mu {\rm m}$\fi}
\def\arcsec{\ifmmode ^{\prime \prime}\else $^{\prime \prime}$\fi}
\def\secpoint{\mbox{$''\mskip-7.6mu.\,$}}

\def\inch{\ifmmode ^{\prime \prime}\else $^{\prime \prime}$\fi}
\def\arcmin{\ifmmode ^{\prime}\else $^{\prime}$\fi}

\def\msun{\ifmmode {\rm M_{\odot}}\else $\rm M_{\odot}$\fi}
\def\mearth{\ifmmode {\rm M_{+\mskip-14.6muO\,}}\else $\rm M_{+\mskip-14.6muO\,}$\fi}
\def\mearth{\ifmmode {\rm M_{\earth}}\else $\rm M_{\earth}$\fi}
\newbox\grsign \setbox\grsign=\hbox{$>$} \newdimen\grdimen \grdimen=\ht\grsign
\newbox\simlessbox \newbox\simgreatbox
\setbox\simgreatbox=\hbox{\raise.5ex\hbox{$>$}\llap
     {\lower.5ex\hbox{$\sim$}}}\ht1=\grdimen\dp1=0pt
\setbox\simlessbox=\hbox{\raise.5ex\hbox{$<$}\llap
     {\lower.5ex\hbox{$\sim$}}}\ht2=\grdimen\dp2=0pt

\begin{document}

\submitted{The Astrophysical Journal Letters, in press}
\title{Forming the Dusty Ring in HR 4796A}
\author{Scott J. Kenyon and Kenneth Wood}
\affil{Smithsonian Astrophysical Observatory, 60 Garden Street, 
Cambridge, MA 02138}
\affil{skenyon@cfa.harvard.edu, kwood@cfa.harvard.edu}
\author{and}
\author{Barbara A. Whitney and Michael J. Wolff}
\affil{Space Science Institute, Suite 23, 1540 30th Street,
Boulder, CO 80303-1012}
\affil{whitney@colorado.edu, wolff@colorado.edu}
\submitted{Received: 7 July 1999, Accepted: 11 August 1999}
%


\begin{abstract}
We describe planetesimal accretion calculations for the dusty 
ring observed in the nearby A0 star HR 4796A.  Models with 
initial masses of 10--20 times the minimum mass solar nebula 
produce a ring of width 7--15 AU and height 0.3--0.6 AU at 70 AU 
in $\sim$ 10 Myr.  The ring has a radial optical depth $\sim$ 1.  
These results agree with limits derived from infrared images and 
from the excess infrared luminosity.

\end{abstract}

\subjectheadings{planetary systems -- solar system: formation -- 
stars: formation -- stars: individual (HR 4796A) -- circumstellar matter}

\section{INTRODUCTION}

HR 4796A is a nearby A star with a large infrared (IR) excess.
Jura (1991) measured the far-IR excess of this wide binary using 
{\it IRAS} data.  Jura \etal (1995, 1998) associated the excess 
with the A0 primary and derived the ratio of the far-IR to 
stellar luminosity, $L_{FIR}/L_{\star} \approx$ 
$ 5 \times 10^{-3}$.  In 1998, two groups reported 
extended thermal emission at $\lambda$ = 20 $\mu$m 
from a dusty disk with an inner hole at $\sim$ 40--70 AU 
assuming the {\it Hipparcos} distance of 67 $\pm$ 3.5 pc
(\cite{jay98}; \cite{koe98}).  Observations with NICMOS aboard 
{\it HST} have revealed a thin annulus of scattered light, with 
a width of $\lesssim $ 17 AU at a distance of $\sim$ 70 AU from 
the central star (\cite{sch99}).  With an age of $\sim$ 10 Myr 
(\cite{sta95}; \cite{bar97}), the A0 star is older than most
pre--main-sequence stars and younger than stars like $\beta$ Pic 
and $\alpha$ Lyr with `debris disks'.

The dusty ring in HR 4796A challenges theories of planet formation.  
In most planetesimal accretion calculations,
planet-sized objects do not form on short timescales at large 
distances from the central star.  Kenyon \& Luu (1999; KL99 
hereafter) estimate formation times of 10--40 Myr for Pluto 
at 35 AU from the Sun.  Achieving shorter timescales at 70 AU in 
HR 4796A requires large initial masses, which might conflict with 
masses derived from IR observations. In the inner Solar System, 
planet formation cannot be confined to a narrow ring, because 
high velocity objects in adjacent annuli interact and `mix' 
planetary growth over a large area (\cite{wei97}).  This problem 
may be reduced at larger distances from the central star, 
where planetary growth is ``calmer''.

Our goal in this paper is to develop planetesimal accretion
models that can lead to the dusty ring observed in HR 4796A.
We begin in \S2 with Monte Carlo calculations to constrain 
the geometry and optical depth of dust in the ring.  In \S3, 
we derive plausible initial conditions which produce the 
observed dust distribution on 10 Myr timescales. These models 
also satisfy constraints on the dust mass from {\it IRAS} 
observations and lead to a self-consistent picture for ring
formation.  We conclude in \S4 with a brief summary and discussion 
of the implications of this study for planet formation in other 
star systems.

\section{MODEL IMAGES}

Current data constrain the geometry and optical depth of the ring.  
Near-IR images measure the amount of scattered light from the 
ratio of the 1.1--1.6 $\mu$m radiation to the stellar luminosity,
$L_{NIR}/L_{\star} \approx$ $2 \times 10^{-3}$ (\cite{sch99}).  
The far-IR luminosity limits the amount of stellar radiation 
absorbed and reradiated.  To construct a physical model, we 
assume an annulus of width $\Delta a$ and height $z$ at a 
distance $a$ = 70 AU from the central star.  The luminosity 
ratios depend on the solid angle 
$\Omega / 4 \pi = 2 \pi a z / 4 \pi a^2 = z/2 a$, 
the radial optical depth $\tau$, and the albedo $\omega$:
$L_{NIR}/L_{\star}$ = $\tau \omega (z / 2 a) $ and
$L_{FIR}/L_{\star}$ = $\tau (1 - \omega) (z/2a)$.
These equations assume gray opacity and scattering 
in the geometric optics limit.
If the annulus contains planetesimals and dust in dynamical 
equilibrium, $z/\Delta a \lesssim$ 1 (\cite{hor85}).  
Anticipating the results of our coagulation calculations, 
where $z/a \sim$ $10^{-2}$, we then have $\omega \approx$ 0.3 --
close to observed values in $\beta$ Pic (\cite{bac93}) -- and 
$\tau \sim$ 1.  

We construct scattered light images using a 3D Monte Carlo code
(\cite{wr99}) with forced first scattering (Witt 1977) and a 
``peeling-off'' procedure (Yusef-Zadeh, Morris, \& White 1984).
We adopt a dust number density, 
$n = n_0 {\rm e}^{-z^2/2 H^2}{\rm e}^{-(a-70)^2/2 A^2}$,
where the scale height $H$ and scale length $A$ are in AU.  
We assume $\omega$ = 0.3, isotropic scattering (see Figure 12 of 
\cite{aug99}), and adjust $\tau$ until the models yield 
$L_{NIR}/L_{\star}$ = $1.5 \times 10^{-3}$ for an input $H$ 
and $A$.  Model images with $\omega \tau$ = constant 
are identical in the optically thin limit.

Figure 1 compares several models with the NICMOS 1.1 $\mu$m image
(from FITS data kindly sent by G. Schneider).  We convolved Monte 
Carlo images with a gaussian point-spread function with FWHM = 
0\secpoint12 to approximate the 0\secpoint12 resolution of NICMOS
(\cite{sch99}).  Model images with $H >$ 5 AU (FWHM = 14 AU) or 
$ A >$ 10 AU (FWHM = 27 AU) are more extended than the data
(\cite{aug99}).  Our preferred model with $H = 0.5$ AU, 
$A = 5$ AU, and $\omega \tau$ = 0.25 reproduces the size and 
shape of the NICMOS image as well as the limb brightening observed 
towards the ring edges.  These results match NICMOS flux ratios best 
for our adopted geometry; larger $H$ implies smaller $\omega \tau$. 
The 3$\sigma$ limit, $\omega \tau$ = 0.12--0.35, agrees with 
previous estimates (cf. \cite{koe98}; \cite{sch99}; \cite{aug99}).  
We disagree, however, with the $\tau \sim 10^{-3}$ 
of Schneider {\it et al.}; their result is valid only for 
scattering in a spherical shell.

\section{COAGULATION MODEL}

To calculate dust evolution in HR 4976A, we use a coagulation code 
based on the particle-in-a-box method (\cite{kl99}).  This 
formalism treats planetesimals as a statistical ensemble of 
bodies with a distribution of horizontal and vertical velocities 
about Keplerian orbits (\cite{saf69}).  We begin with a size 
distribution of $N_i$ bodies having total mass $M_i$ in each 
of $i$ mass batches.  Collisions among these bodies produce 
(i) growth through mergers along with cratering debris for low impact 
velocities or (ii) catastrophic disruption into numerous small 
fragments for high impact velocities.  Inelastic collisions, 
long range gravitational interactions (dynamical friction and 
viscous stirring), and gas drag change the velocities of the mass 
batches with time.  The code has been tested against analytic 
solutions of the coagulation equation and published calculations 
of planetesimal growth.  Although inappropriate for the last 
stages of planet formation, our approach well-approximates the 
early stages (\cite{kok96}).

We model planetesimal growth in an annulus of width $\Delta a$ 
= 12 AU centered at $a$ = 70 AU. The central star has a mass of 
2.5 \msun. The input size distribution has equal mass in each
of 38 mass batches with initial radii $r_i$ = 1--80 m.  For a 
Minimum Mass Solar Nebula with mass $M_{MMSN}$, the total mass
in the annulus is $M_0 \approx$ 15 $M_E$; the initial number 
of bodies with $r_i$ = 1 m is $N_0 \approx 3 \times 10^{20}$.
All batches start with the same initial velocity.  The mass 
density $\rho_0$ = 1.5 g cm$^{-3}$, intrinsic strength 
$S_0 = $ $2 \times 10^6$ erg g$^{-1}$, and other bulk properties 
of the grains are adopted from earlier work (see KL99).

Planetesimal growth at 70 AU follows the evolution described previously 
(KL99).  The 80 m bodies first grow slowly into 1~km objects.  During 
this slow growth phase, frequent collisions damp the velocity dispersion 
of all bodies.  ``Runaway growth'' begins when the gravitational range 
of large objects exceeds their geometric cross-section.  These bodies 
grow from 1~km up to $\sim$ 100 km in several Myr.  During runaway 
growth, collisional debris, dynamical friction, and viscous stirring 
increase the velocity dispersion of small bodies from 
$\sim$ 1~m~s$^{-1}$ up to $\sim$ 40~m~s$^{-1}$.  This evolution 
reduces the gravitational range of the 100~km objects and ends 
runaway growth.  The largest objects then grow slowly to 1000+ km 
sizes.

Figure 2(a) shows the growth of the largest object in several models.
For $M_0$ = 10 $M_{MMSN}$ and $e_0$ = $10^{-3}$, Pluto-sized objects 
form in $t_P$ = 2.1 Myr at $a$ = 35 AU, 13 Myr at 70 AU, and 
93 Myr at 140 AU. Models with smaller $M_0$ take longer to 
make ``Pluto''.  Plutos form more quickly for $e_0 < 10^{-3}$,
because gravitational focusing factors are larger.

Figure 2(b) shows the evolution of the scale height $H$ for small 
objects.  Initially, $H = 2 \pi a~{\rm sin}~i \le 0.003 a$ 
for $e_0 \le 10^{-3}$.  Collisional damping cools the bodies 
during the slow-growth phase; $H$ remains small.  $H$ increases
dramatically during runaway growth, when dynamical processes heat up
the smallest bodies.  Once runaway growth ends, $H$ slowly increases
to 0.3--0.6 AU independent of $M_0$, $e_0$, and other input parameters.

When $H$ begins to increase, high velocity collisions produce
numerous ``dust grains'' with sizes $\lesssim$ 1 m.  We do not
follow explicitly the evolution of these bodies.  Instead, we 
assume that collisional debris is
(i) swept up by 1 m or larger objects,
(ii) ejected by radiation pressure, or
(iii) dragged inwards by the Poynting-Robertson effect.
Grains with sizes exceeding 4--5 $\mu$m are stable against 
radiation pressure (\cite{jur98}; \cite{aug99}).  
Poynting-Robertson drag reduces the mass in small grains on a 
timescale $t_{PR} \approx$ 1.0 Myr ($r_i / 4~\mu$m).  With the 
short collision times, $\le 10^5$ yr, in our model annulus, 
1 Myr seems a reasonable estimate of the timescale for collisions 
to produce 4 $\mu$m grains which are removed by radiative processes. 
For this paper, we calculate the accretion explicitly and 
adopt a 1 Myr timescale for dust removal.  

Figure 2(c) shows the dust mass as a function of time.
The results are not sensitive to the adopted mass distribution 
for grains with $r_i \gtrsim$ 4 $\mu$m or to factor of 2--3 
variations in the removal timescale.  The dust mass is initially 
large due to the starting conditions.  The dust mass decreases 
with time, because 
(i) collisional damping of the smaller bodies leads to less 
collisional debris and (ii) radiative processes and accretion
by large bodies remove dust.
Once runaway growth begins, collisions between small bodies 
produce more dust.  The dust mass then reaches a rough 
equilibrium between collision debris and dust removed by 
radiation forces and by the larger bodies.

These results indicate that large dust masses correlate with runaway 
growth and the formation of 1 or more Plutos in the outer parts 
of the disk.  To predict the amount of radiation absorbed and 
scattered by dust and larger bodies, we compute $\tau$ from 
the model size distribution.  We assume the geometric optics 
limit because $r_i \gg \lambda$.  For the large bodies 
$\tau = \sum_{i=1}^N n_i \sigma_i \Delta a$, 
where $n_i$ is the number density in mass batch $i$,
$\sigma_i$ is the extinction cross-section, and 
$N$ is the number of mass batches.  
We adopt $\sigma_i = 2 \pi r_i^2$ and a volume
$V_i = 2 \pi a \Delta a H_i$ to compute $n_i = N_i/V_i$
and hence $\tau$ for material with $r_i \gtrsim$ 1 m.

Estimating $\tau$ for small particles requires an adopted 
cumulative size distribution, $N_C \propto r_i^{-q}$.  
We consider three choices:
(i) $q = 2.5$, the collisional limit for coagulation;
(ii) $q = 3$, equal mass per mass interval; and
(iii) $q = 3.5$, the approximate distribution for
grains in the interstellar medium.  Our calculations produce
$q \approx 2.7$ for 1--100 m bodies.  We expect a slightly steeper 
mass distribution for smaller bodies, because collisions between 
smaller bodies produce fewer mergers and more debris.

Figure 2(d) shows how $\tau$ evolves for a single model.  
The large bodies initially have modest radial optical depth, 
$\tau_L \approx$ 0.2.  This optical depth decreases with time, 
except for a brief period when runaway growth produces 10--100 km
objects with small scale height above the disk midplane.
The large bodies are transparent once a Pluto forms.  
The small grains are also initially opaque.  This dust is 
transparent at late times if most of the mass is in the 
largest grains, $q \lesssim$ 2.8.  The dust is opaque for 
$q \gtrsim$ 3.

Table 1 summarizes results for various initial conditions.  
Models with $M_0 \approx$ 10--20 $M_{MMSN}$ and 
$e_0 \approx$ $10^{-4}$--$10^{-3}$ achieve $\tau \sim 1$ in 
10 Myr.  Less massive disks produce less dust on longer timescales.
The results are not sensitive to other input parameters, including 
the size distribution and the bulk properties of the bodies.

Table 1 also shows why dust in HR 4796A lies in a ring. 
In disks with surface density $\Sigma \propto a^{-3/2}$, the Pluto 
formation timescale\footnote{Pluto is a handy reference: 1000+ km 
objects form roughly in the middle of the rapid increase in $H$ 
which produces large dust masses.} is $t_P \approx$ 
13 Myr $(M_0/10~M_{MMSN})^{-1}$ $(a/{\rm 70~AU})^{2.7}$.  
Once an annulus at $a$ begins to form dust, material at 
$a + \Delta a$ must wait a time, $\Delta t/t_p \approx 2.7 \Delta a/a$, 
to reach the same state.  This result sets a hard outer limit 
to the ring, $\Delta a /a \approx 0.4 \Delta t/t_p \approx$ 
0.1--0.2, if $\Delta t$ is the time for $H$ to double 
in size during runaway growth, 2--3 Myr.  We expect a hard inner
edge, because particle velocities reach the shattering limit of
$\sim$ 100 m s$^{-1}$ (KL99) or planets sweep up the dust 
(e.g., \cite{pol96}) or both.

\section{DISCUSSION AND SUMMARY}

Our results indicate that the dusty ring in HR 4796A is a natural
outcome of planetesimal evolution.  Planet formation at 70 AU in 
10 Myr is possible with an initial disk mass of 10--20 $M_{MMSN}$.
Dust production associated with planet formation is then confined 
to a ring with $\Delta a \approx$ 7--15 AU. The optical depth in 
this ring satisfies current constraints on scattered light at 
1--2 $\mu$m and on thermal emission at 10--100 $\mu$m if the size 
distribution of the dust is $N_C \propto r_i^{-q}$ with 
$ q \gtrsim 3$ for $r_i \lesssim$ 1 m.
Models with disk masses smaller than 10 $M_{MMSN}$ fail to 
produce planets and an observable dusty ring in 10 Myr.

An uncertainty in our model is the timescale to produce 
1--80 m bodies from small dust grains in a turbulent, gaseous disk.
Cuzzi \etal (1993) show that grains grow very rapidly once 
they decouple from eddies in the disk.  The decoupling 
timescale depends on the unknown disk viscosity at 70 AU.

Our model makes several observational predictions.  We expect
$L_{NIR}/L_{\star}$ = constant for $\lambda \le$ 5 $\mu$m; current 
data are consistent with this prediction at the 1.5$\sigma$ level. 
Better measurements of the ring flux at $\lambda \ge$ 1.6 $\mu$m
would test our optical depth assumptions and yield interesting 
constraints on grain properties.
Deep images at $\lambda \ge$ 10--20 $\mu$m with high spatial 
resolution should detect material outside the ring.  We predict 
$\tau \approx$ 0.1 in large bodies for $a \gtrsim$ 80 AU;
the surface brightness and temperature of this material should 
decrease markedly with radius.  This material should have negligible
mass in small objects, because coagulation concentrates mass in
the largest objects when $H$ is small.  We also expect a flux of 
dust grains into the central star, although we cannot yet compare 
quantitative predictions with observations.  Future calculations 
of radiative processes within the ring will address this issue.

Applying this HR 4796A model to other stars with circumstellar disks 
is challenging due to small number statistics and unfavorable
circumstances.  Nearby companion stars probably influence the 
dynamics of dusty rings in HD 98800 and HD 141569
(\cite{pir97}; \cite{low99}; \cite{lag00}).  In HR 4796A, the 
M-type companion lies well outside the ring radius and cannot 
modify ring dynamics significantly.  Older systems like 
$\beta$ Pic and $\alpha$ Lyr require time-dependent treatment of 
dust to allow the ring to spread with time (e.g., \cite{art97}).  
We plan to incorporate this time-dependent behavior in future 
calculations to see whether the ring in HR 4796A can evolve into 
a debris disk (as in e.g. $\beta$ Pic and $\alpha$ Lyr) on a timescale
of 100--200 Myr.

The main alternative to {\it in situ} ring formation at 70 AU is 
migration of a planet formed at a smaller radius.  Weidenschilling
\& Marzari (1996) show that gravitational interactions can scatter 
large objects into the outer disk in less than 1 Myr.  Migration 
reduces the required ring mass by a factor of 10--100.  However, 
the scattered body has a large eccentricity, $e \sim 0.5$.  
Dynamical friction might circularize the orbit in 10 Myr, but 
would induce large eccentricities in smaller bodies. The width 
of the dusty ring would probably exceed observational constraints.  
Future calculations can address these issues.

We thank B. Bromley for helping us run our code on the HP Exemplar 
``Neptune'' at JPL and for a generous allotment of computer time 
through funding from the NASA Offices of Mission to Planet Earth, 
Aeronautics, and Space Science.

\vfill
\eject


\hskip 5ex
\epsfxsize=2.1in
\epsffile{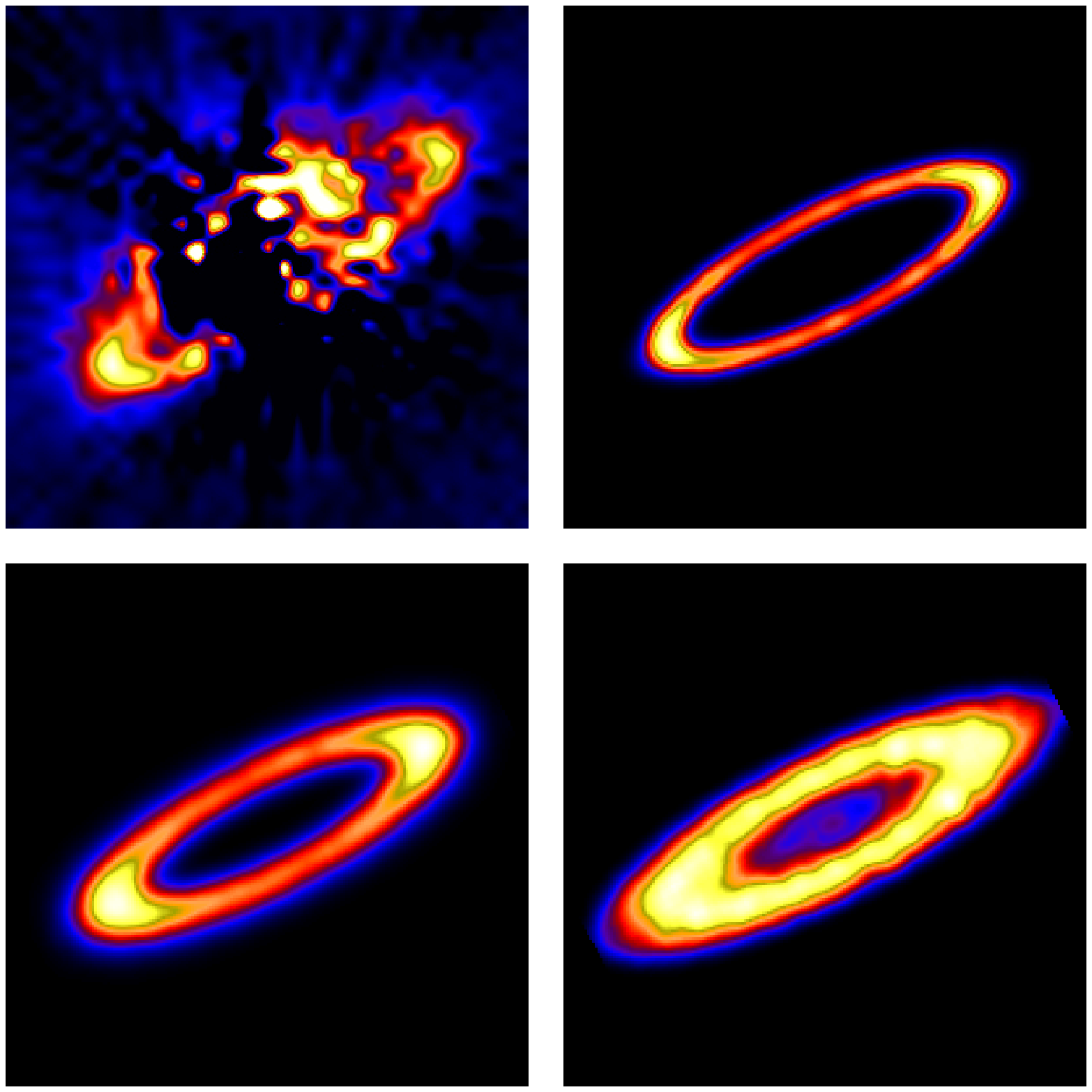}
\vskip 40ex
\figcaption{Comparison of model scattered light images with HST data
of HR 4796A.
(a) upper left panel: NICMOS coronagraphic image at 1.1 $\mu$m.
(b) upper right panel: Model scattered light image with 
$z$ = 0.5 AU, $R$ = 5 AU, and $\omega \tau_{NIR}$ = 0.25.
(c) lower left panel: As in (b) for $z$ = 5 AU, $R$ = 10 AU, 
and $\omega \tau_{NIR}$ = 0.02.
(d) lower right panel: As in (c) for $z$ = 1 AU, $R$ = 20 AU, 
and $\omega \tau_{NIR}$ = 0.1.} 

\epsfxsize=7.0in
\epsffile{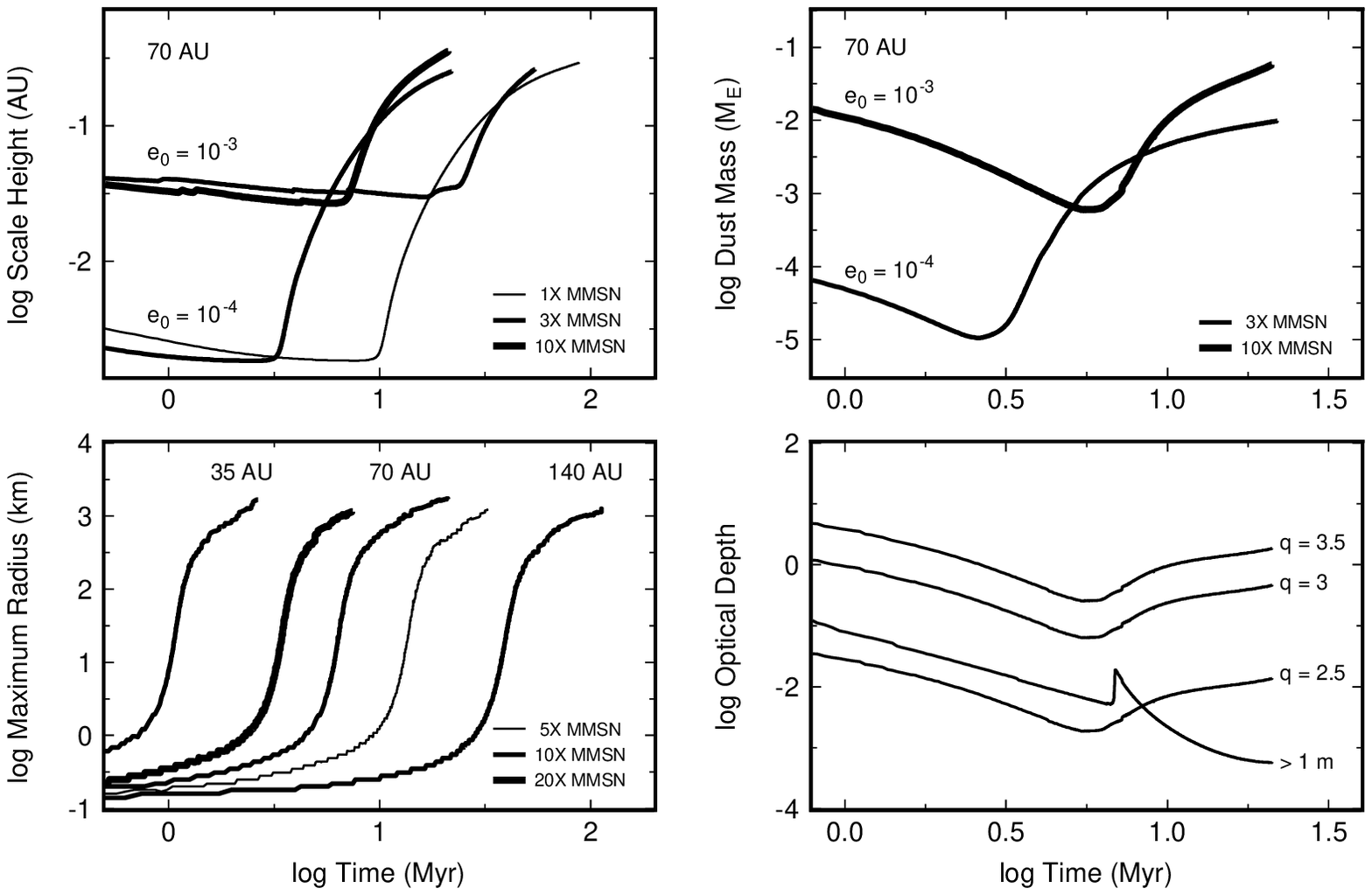}
\vskip -12ex
\figcaption{Results for HR 4796A coagulation models.
(a) lower left panel: maximum radius for models with $e_0 = 10^{-3}$
at $a$ = 35, 70, and 140 AU; initial masses are listed in the legend.
(b) upper left panel: scale height of small bodies for 70 AU models
with $e_0$ and $M_0$ as listed.
(c) upper right panel: dust mass for 70 AU models with 
$e_0$ and $M_0$ as listed.
(d) lower right panel: optical depth for a 70 AU model with 
$e_0 = 10^{-3}$ and $M_0 = 10 M_{MMSN}$.}

\clearpage

\begin{center}
\begin{deluxetable}{l r c r c c}
\singlespace
\tablewidth{15cm}
\tablenum{1}
\tablecaption{Results of Planetesimal Accretion Calculations}
\tablehead{
\colhead{$a$} & \colhead{$M_0$} & 
\colhead{~~~~$e_0$~~~~} & \colhead{$t_P$} & 
\colhead{log $\tau_S$} & \colhead{log $\tau_L$}}
\startdata
35 & 100  & $10^{-3}$ & 2.1~~~ & ~~0.09 & $-$3.43 \\
\\
70 & 15  & $10^{-4}$ & 81.1~~~ & $-$1.90 & $-$3.92 \\
70 & 45  & $10^{-4}$ & 20.4~~~ & $-$0.96 & $-$3.27 \\
70 & 150 & $10^{-4}$ &  5.7~~~ & $-$0.04 & $-$2.69 \\
70 & 15  & $10^{-3}$ &156.4~~~ & $-$2.21 & $-$4.24 \\
70 & 45  & $10^{-3}$ & 50.0~~~ & $-$1.42 & $-$3.65 \\
70 & 75  & $10^{-3}$ & 29.9~~~ & $-$0.94 & $-$3.48 \\
70 & 150 & $10^{-3}$ & 13.0~~~ & $-$0.34 & $-$3.25 \\
70 & 300 & $10^{-3}$ &  6.6~~~ & $-$0.05 & $-$2.62 \\
\\
140 & 200 & $10^{-3}$ & 92.6~~~ & $-$1.20 & $-$2.95 \\
\enddata
\tablecomments{$a$ is the distance of the annulus from
the central star in AU; $M_0$ is the initial mass in the
annulus in $M_E$; $e_0$ is the initial eccentricity of 
each mass batch; $t_P$ is the timescale in Myr to produce
Pluto-sized objects; $\tau_S$ is the optical depth in dust 
when the first Pluto forms, assuming equal mass in dust
per decade in radius; $\tau_L$ is the optical depth of 
the large bodies when the first Pluto forms.}
\end{deluxetable}
\end{center}

\vfill
\eject





\end{document}